# Alternative interpretation of cosmic ray investigations above the knee

A. G. Bogdanov, R. P. Kokoulin, A. A. Petrukhin
*National Research Nuclear University MEPhI (Moscow Engineering Physics Institute), Moscow, 115409, Russia*

All information about primary cosmic rays above the knee has been obtained from results of EAS investigations. At that, two alternative approaches exist: cosmophysical and nuclear physical. In the frame of the first one, all changes in measured EAS characteristics are explained by the changes in energy spectrum and mass composition of primary cosmic rays. In this paper, the second approach is considered, in frame of which corresponding changes in EAS parameters are explained by changes of interaction model above the knee. Some experimental possibilities of proof of the correctness of the nuclear physical approach are considered.

## 1. INTRODUCTION

In general, cosmic ray energy spectrum can be described by a power function $E^{-\gamma}$ with $\gamma$ ~ 2.7 [1]. But there are various deviations from this simple function: the knee, the second knee, the ankle, etc. All these features appear above the energy $10^{15}$ eV. And namely this energy is the limit for direct measurements of primary CR energy spectrum and composition at satellite and balloon experiments. Above $10^{15}$ eV, investigations of extensive air showers (EAS) are possible only, and all so-called experimental data about CR energy spectrum and composition are results of such investigations. However, measured characteristics of EAS ($N_e$, $N_\mu$, $E_{core}$, $X_{max}$, etc.) depend not only on primary particle type and its energy, but on the model of interaction, too. Now all measured changes in EAS characteristics in dependence on EAS size are explained by changes of CR energy spectrum and composition. However, in spite of numerous attempts, no exhaustive and generally accepted explanation of these changes was proposed.

In this paper, an alternative approach is considered in which primary energy spectrum and composition are practically not changed, and all observed changes in EAS characteristics beginning from the first knee in the spectrum of the EAS in the number of charged particles are explained by the change of interaction model. Additional stimuls to consider such approach are various unusual events (Halos, Alignment, Penetrating cascades, Centauros and others [2]) which were observed in different experiments namely at energies above $10^{15}$ eV. And the last argument in favor of interaction model change is the so-called “muon puzzle” [3] which includes two types of experimental data: 1) excess of muons in EAS compared to calculations performed using various existing interaction models, even for extremely heavy mass composition of cosmic rays (pure iron nuclei) [4, 5] and 2) excess of very high energy (VHE) muons with energy > 100 TeV [6].

## 2. SHORT DESCRIPTION OF A POSSIBLE NEW INTERACTION MODEL

Ideas and main aspects of a new interaction model were discussed in many papers (see f.e. [7, 8]). Therefore here only a short description of this model is given. But before introducing such description, it is necessary to answer the important question: what requirements must satisfy a new interaction model?

To explain all unusual events and phenomena observed in cosmic rays, including various changes of their energy spectrum and composition, a model of nucleus-nucleus interaction must provide: threshold behavior at energy about the knee; large cross-section, which must be sufficient for detection in cosmic ray experiments; large orbital momentum for explanation of the alignment and other unusual phenomena observed in CR; large yield of muons and VHE muons to explain both parts of muon puzzle; change of EAS development and their main characteristics: total number of charged particles, the $N_\mu/N_e$ ratio, $X_{max}$ position, etc.

To satisfy these requirements, a model of a blob of quark-gluon matter with a large orbital momentum which is produced in nucleus-nucleus interactions, since cosmic rays even at normal composition consist mainly (~ 60%) of nuclei which interact with nuclei of atoms of the atmosphere, may be considered. Firstly, this model provides a threshold behavior, since for the QGM production a high temperature (energy) is required. Secondly, it provides a large cross-section since for QGM production a transition from quark-quark interaction to interaction of many quarks occurs. In this case geometrical cross-section for point particles $\sigma = \pi\lambda^2$ is changed and will be of the order of $\sigma = \pi R^2$, were $R$ is an effective size of QGM blob.

The next requirement is a large orbital momentum. A possibility of its appearance in non-central ion-ion collisions was considered in paper [9]. It was shown that the value of the orbital momentum is proportional to $\sqrt{s}$ and can be relatively large. Further investigations [10] showed that its value can achieve ~ $10^4$ in collisions of heavy nuclei (f.e. Au-Au).

The main idea of considered model is the existence of some resonance state of QGM blob with mass about 1 TeV and large orbital momentum, which gives correspondingly large centrifugal barrier

$$V(L) = \frac{L^2}{2mR^2}. \quad (1)$$

Such dependence on particle mass changes the picture of QGM blob decay. Strong suppression of decays into light quarks gives a time for production of heavy quark-antiquark pairs, including $t$-quarks, for which centrifugal barrier will be $m_t/m_u \sim 10^5$ times less than for $u$-quarks.

Appearance of $t$-quarks changes EAS development and characteristics of various components of secondary cosmic rays in the atmosphere. Top-quarks very rapidly (~ $10^{-25}$ s) decay into $b$-quark and $W$-boson

$$t(\bar{t}) \to b(\bar{b}) + W^+(W^-)\,, \qquad (2)$$

$b$-quark can produce a jet, in which transitions $b \to c \to s \to d(u)$ are possible. $W$-boson decays into hadrons (mainly pions, on average ~ 20) with probability ~ 66% and leptons (~ 34%).

In the first case, multiplicity of secondary particles is increased in comparison with predictions of existing models, in the second case an excess of VHE muons and neutrinos appears. Three neutrinos and VHE muons increase sharply a missing energy, and EAS energy $E_2$ will be not equal to primary particle energy $E_1$.

## 3. CHARACTERISTICS OF SECONDARY COSMIC RAYS IN THE FRAME OF THE NEW MODEL

Production of QGM blobs by primary CRs in the atmosphere leads to the following consequences.

1. Change of the EAS cascade curve. In Figure 1, results of EAS cascade curve simulation are presented. It is seen that an inclusion of $t\bar{t}$ - pair production in cascade generated by proton gives cascade curve close to cascade generated by iron nucleus in frame of usual model, due to the increase of secondary particle number from $W$-decays. A remarkable decreasing of cascade energy (proportional to the area under the cascade curve) gives missing energy and changes the energy spectrum of detected EAS.

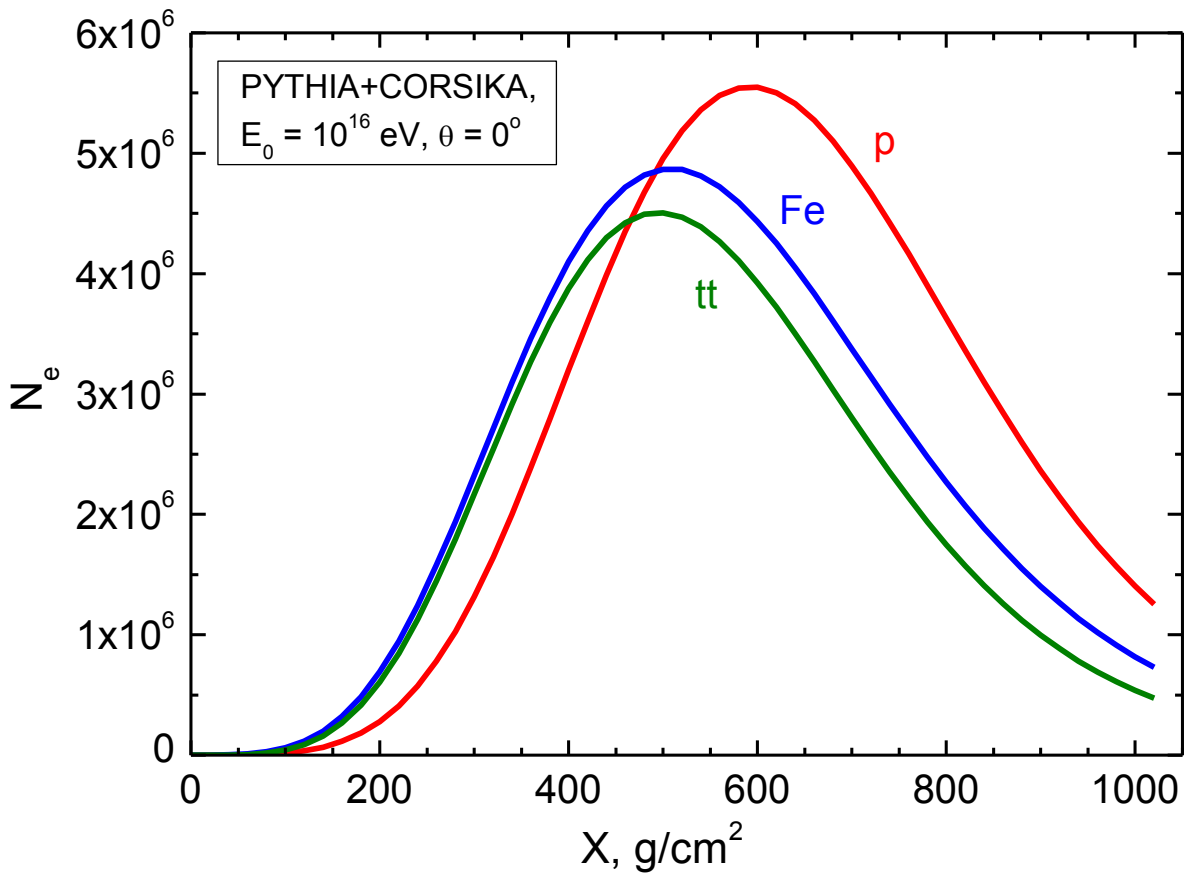

Figure 1. EAS cascade curves in the atmosphere for p, Fe primary particles in the frame of standard interaction model and for protons with production of $t$-quarks.

2. Excess of VHE muons. Decays of $W$-bosons into leptons ($e\nu_e$, $\mu\nu_\mu$, $\tau\nu_\tau$) lead to their excess compared to conventional energy spectra. In Figure 2, the results of calculations of differential muon energy spectrum are presented. Similar energy spectra will be for three types of neutrinos. Since muon energies are not measured and neutrinos are not registered, this gives a large value of missing energy. That leads to large difference between primary energy $E_1$ and EAS energy $E_2$.

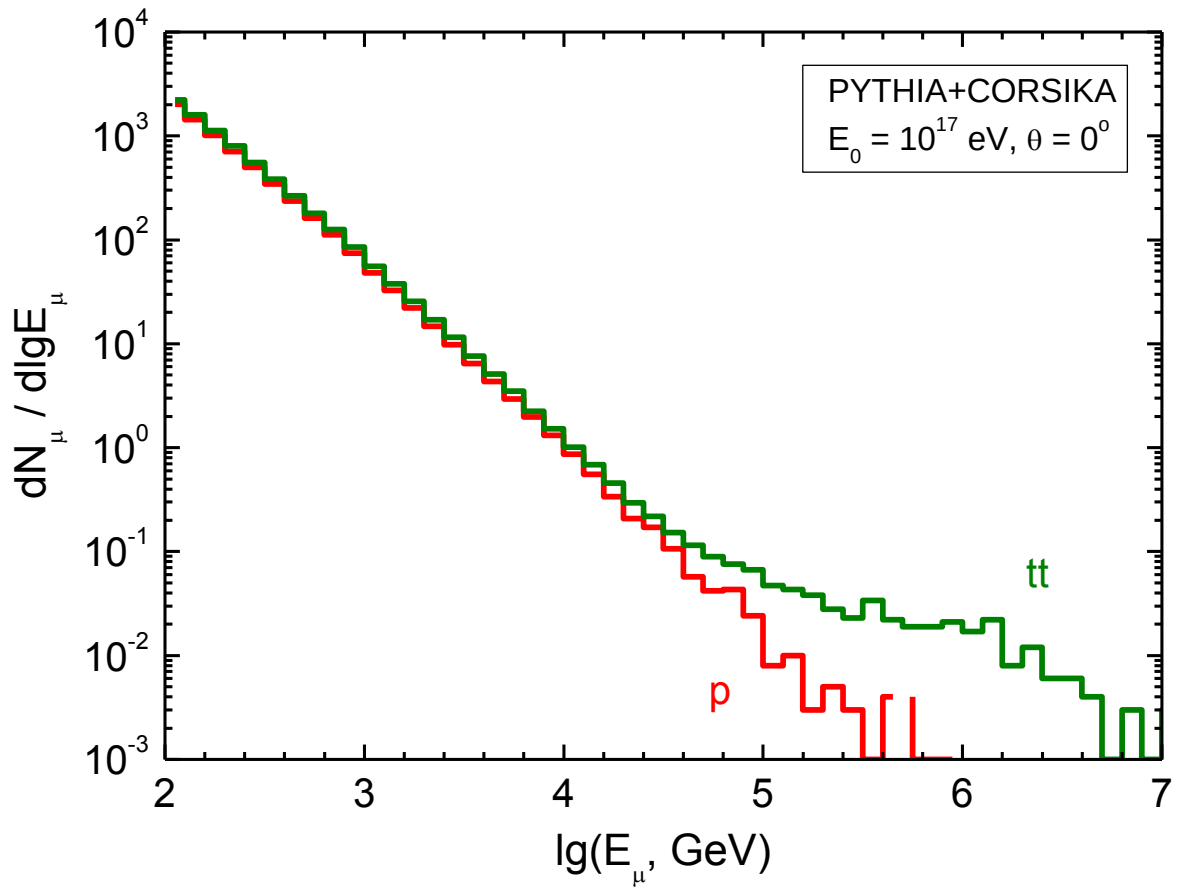

Figure 2. The inclusive muon energy spectrum simulated by means of CORSIKA taking into account $t$-quarks pair production according to PYTHYA.

3. Change of the measured energy spectrum. Appearance of a large missing energy will change cosmic ray energy spectrum, more exactly, the spectrum of EAS in the number of charged particles. If to not take into account this change of interaction model and to use existing models for transition from EAS energy to energy of cosmic ray particles, a change of CR energy spectrum will be observed. Figure 3a illustrates the situation for sharp threshold energy. If to not take into account a difference between $E_1$ and $E_2$, some bump and the knee in energy spectrum will be obtained (Figure 3b).

4. Change of CR composition. Transition from quark-quark interaction to interactions of many quarks and gluons changes the energy in the center-of-mass system of interacting nuclei, too. Usual formula $\sqrt{s} = \sqrt{2m_N E_1}$ must be changed to $\sqrt{s} = \sqrt{2m_c E_1}$, were $m_c$ is some compound mass, which in the first approximation can be written as $m_c = nm_N$ with $1 \le n \le A$, an effective mass of interacting nuclei. It means that $\sqrt{s}$ is abruptly increased, and correspondingly is increased the orbital momentum. If these new values $\sqrt{s}$ and $L$ coincide with parameters of some resonance state (blob) of quark-gluon matter, it will be possible to observe results of decay of this QGM blob into heavy quarks.

Since for production of a new state of matter (QGM blob) not only high temperature (energy), but also a high density of matter is required, the threshold energy for its production will be less for heavy nuclei than for light nuclei and protons. Therefore, in contrast with traditional approach in cosmic rays, firstly iron energy spectrum begins to change, then spectra of more light nuclei, and the last the spectrum of protons. It is important to mark that the measured spectra of various nuclei will be not equal to primary spectra! Results of respective calculations are shown in Figure 4.

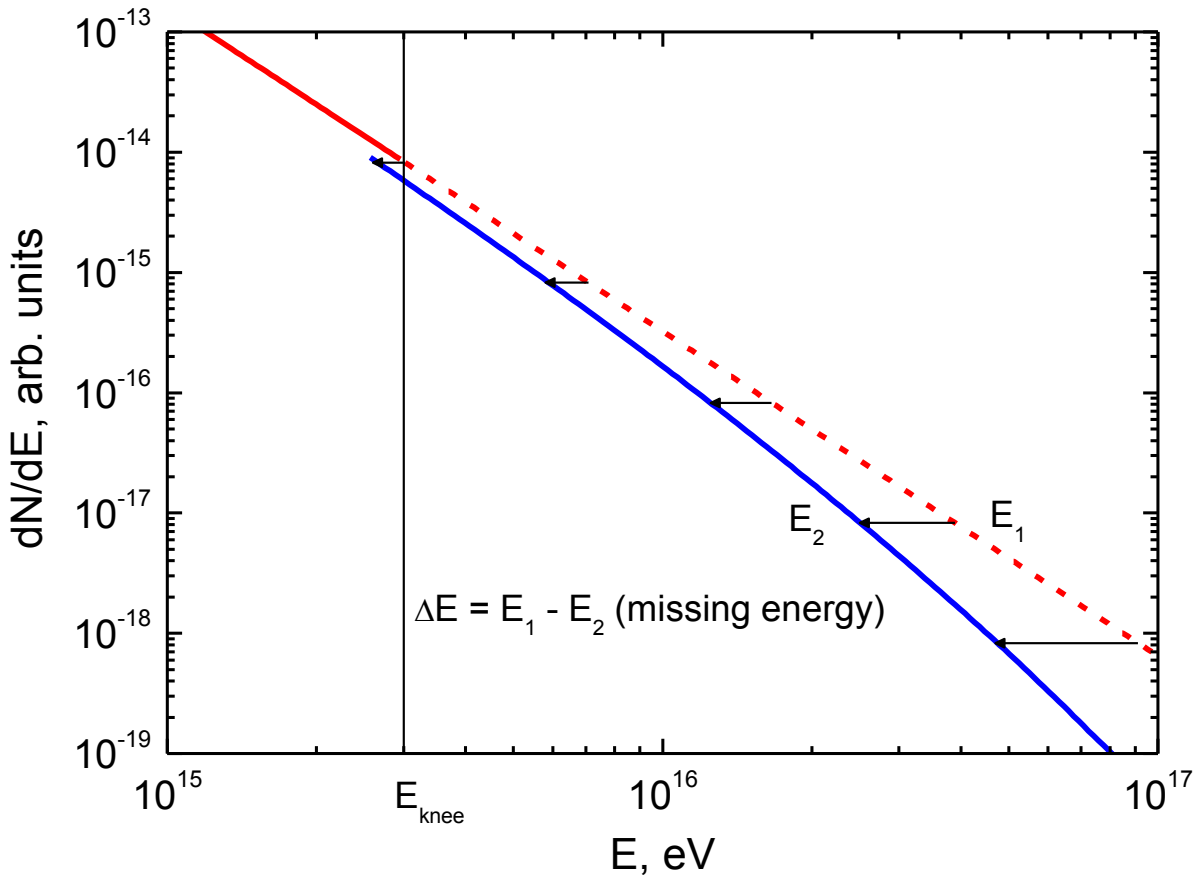

Figure 3,a. The change in the CR energy spectrum at the appearance of the missing energy.

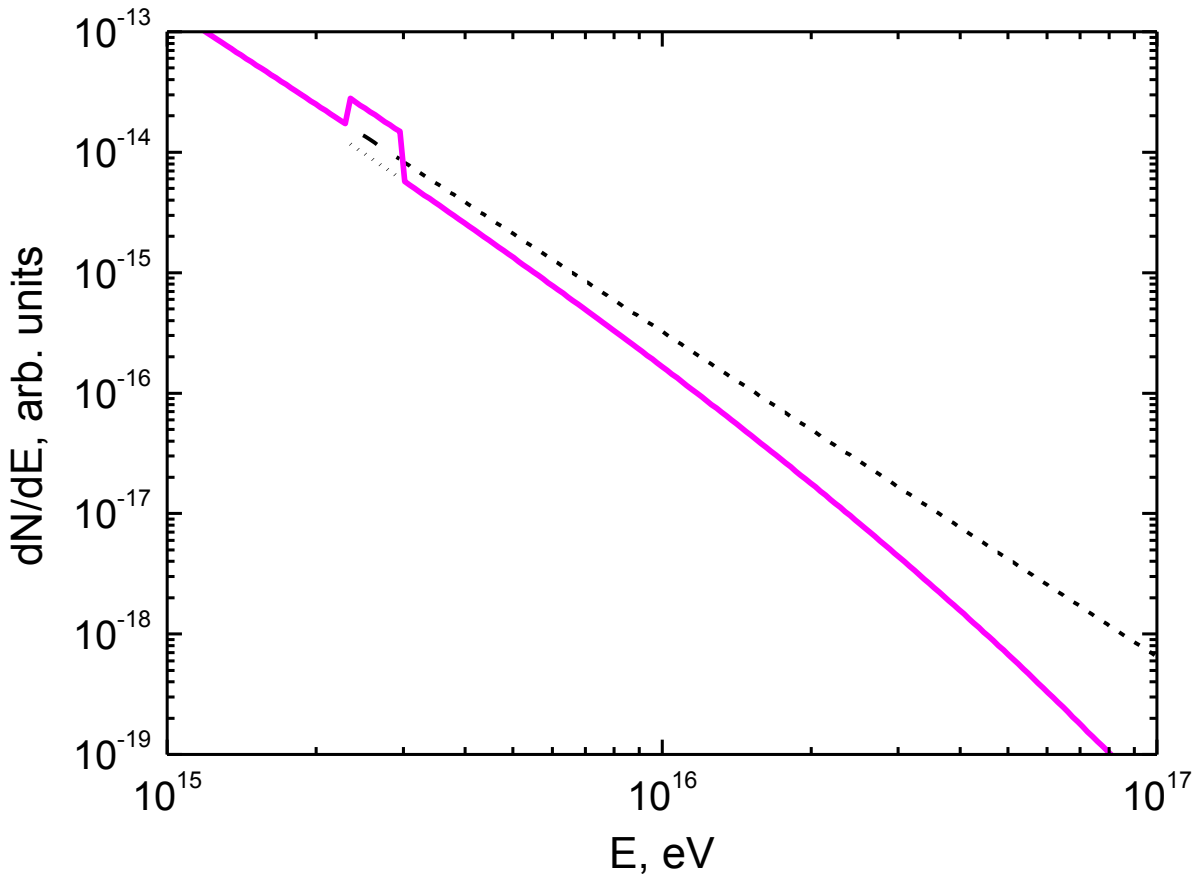

Figure 3,b. The production of the knee with some "bump" in the nuclear-physical approach.

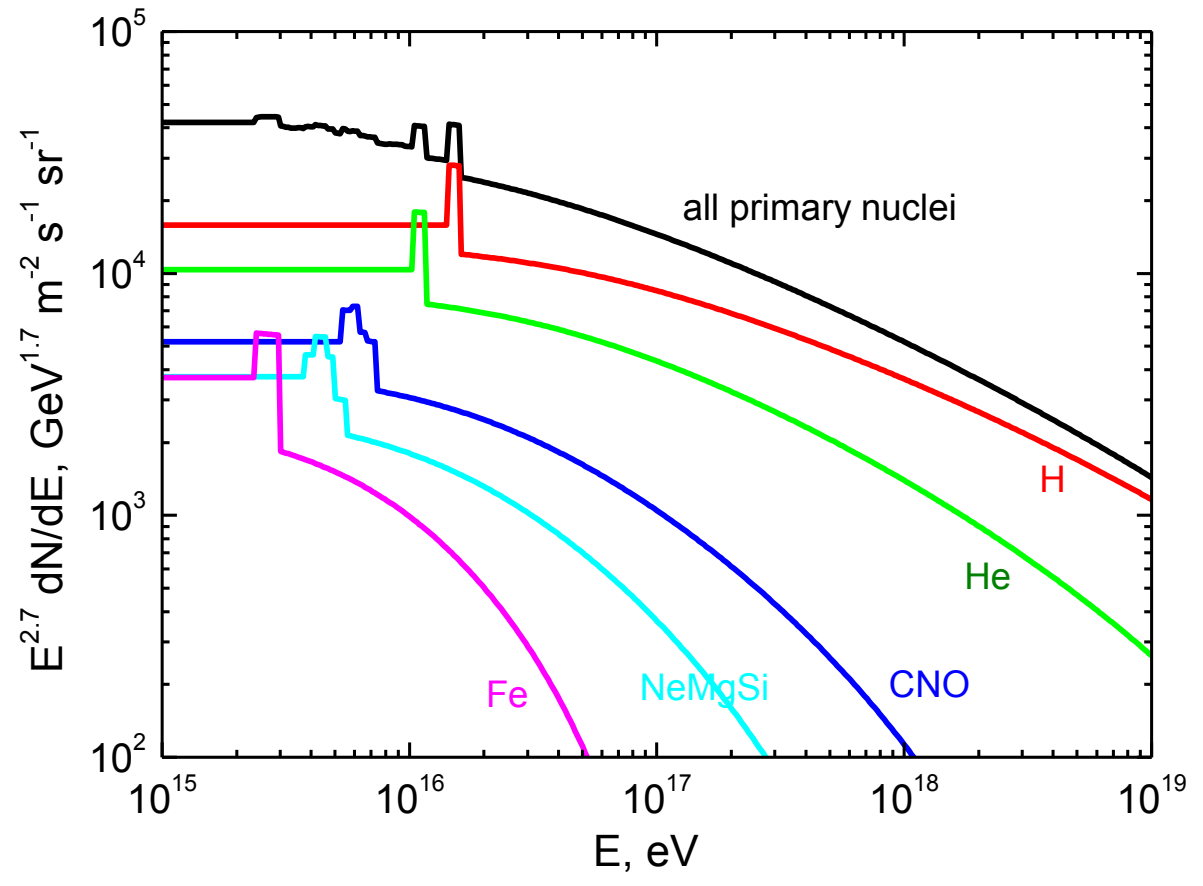

Figure 4. Changes of various CR nuclei spectra in the frame of the considered interaction model.

The all-particle spectrum obtained in the frame of the new model is in a good agreement with experimental data (Figure 5). At that, the measured composition will be enriched by heavy nuclei though primary CR composition is not changed compared to a standard composition at moderate energies.

So, the proposed approach shows a good agreement with experimental data, but their interpretation is principally different. Changes are connected not with changes of primary cosmic ray energy spectrum and composition, but with changes of the interaction model.

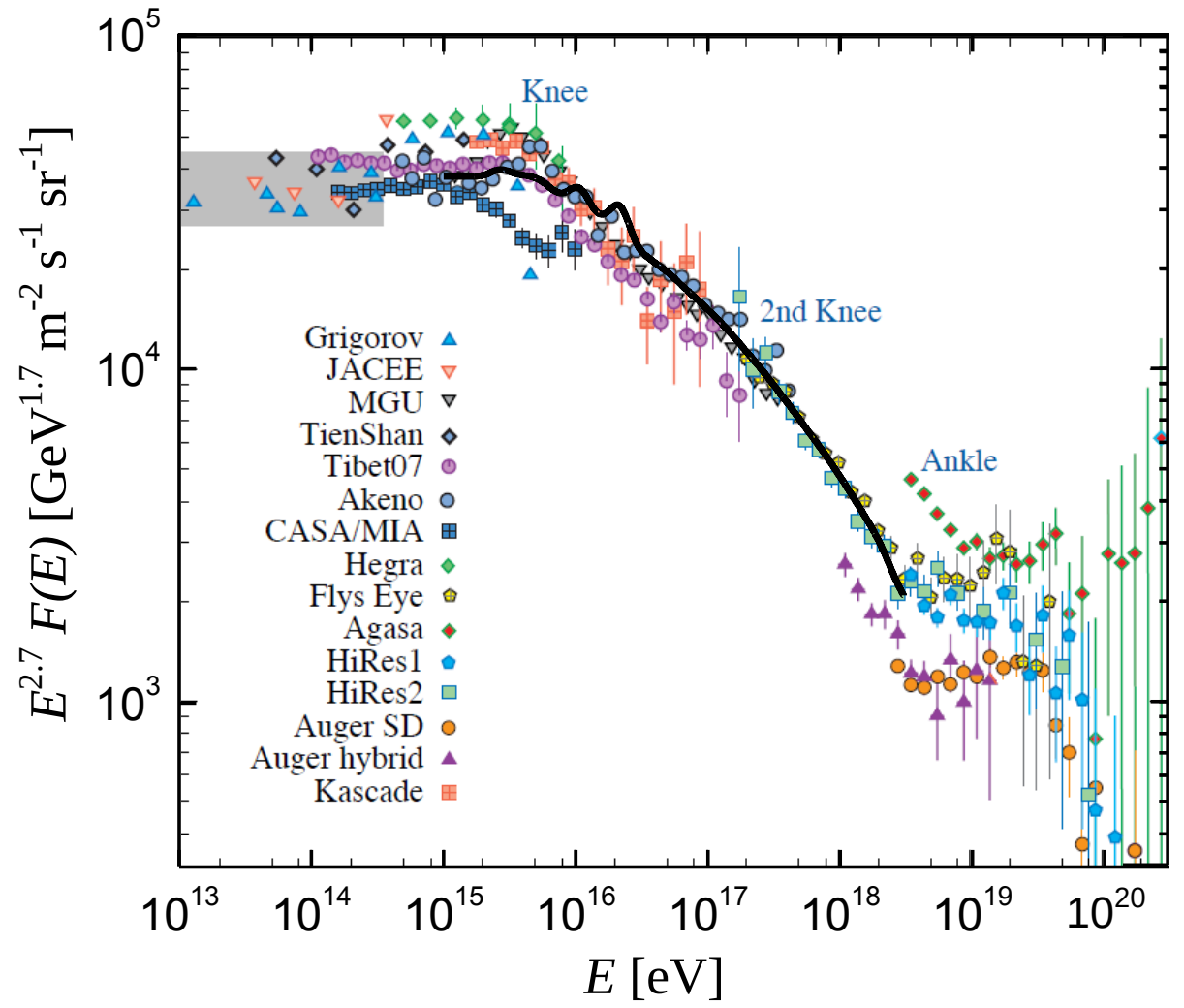

Figure 5. All-particle spectrum calculated in the frame of the new interaction model and experimental data.

## 4. HOW TO CHECK THE NEW INTERACTION MODEL?

Corresponding possibilities exist both in cosmic ray investigations and in LHC experiments.

In cosmic rays, most reliable results can give measurements of energy characteristics of muons. The observation of an excess of VHE (> 100 TeV) muons in inclusive energy spectrum will be an irrefutable proof of heavy particle production which decay into muons (and neutrinos, of course). The first result of such excess was obtained in [6]. Its confirmation is possible in more appropriate detectors, f.e. IceCube or other neutrino water detectors, which can be used for muon energy measurements by using pair-meter technique [11].

Another possibility is connected with the search of high energy muon bundles. In this case it is not possible to measure energy of separate muons, and only energy deposit of detected muon bundle can be measured. For that, two types of detectors are required: one for muon number measurement and the second for energy deposit measurement. These measurements are performed in NEVOD-DECOR experiment in which coordinate-tracking detector and Cherenkov water calorimeter are combined [12].

Various possibilities to check the new model are available in LHC experiments. Predicted by the new model effects – excess of *t*-quarks, excess of *W*-bosons, a sharp increase of missing energy – can be observed by existing LHC detectors. But for that A-A-interactions must be investigated, since in this case the corresponding cross-sections will be larger. Apparently some observations of the effects predicted by the new model were done in Pb-Pb-interactions [13] in which a fast increase of secondary particle multiplicity was measured. The last experiments give some excess of *W*-bosons even in p-p-interactions [14-16].

## 5. CONCLUSION

Usually, for interpretation of results of EAS characteristics measurements so-called cosmo-physical approach is used. According to this approach all changes in EAS characteristics are explained by changes of primary cosmic ray energy spectrum and composition. As it is shown in this paper, these observed changes in EAS characteristics can be explained by the change of the interaction model. The considered interaction model is based on the threshold production of QGM blobs with a large orbital momentum. This model explains all detected changes in EAS characteristics and additionally practically all unusual events observed in various experiments at energies above the knee [17]. Besides that, the new model explains also the “muon puzzle” [3].

## Acknowledgments

This work was performed with the support of the Ministry of Education and Science of the Russian Federation (government task and MEPhI Academic Excellence Project 02.a03.21.0005).